\documentclass[pra,aps,twocolumn,showpacs,epsfig]{revtex4}
\usepackage{color}
\usepackage{graphicx,amssymb,epsfig,epsf,amsmath}

\begin{document}
\title{Macroscopic quantum many-body tunneling of attractive Bose-Einstein condensate 
in anharmonic trap}

\author{
Sudip Kumar Haldar$^{1}$\footnote{e-mail: sudip\_cu@rediffmail.com}, Pankaj Kumar Debnath$^{2}$, and Barnali Chakrabarti$^{3}$
}


\affiliation{
$^{1}$Department of Physics, Lady Brabourne College, P-1/2 Surawardi Avenue, Kolkata-17, India.\\
$^{2}$Santoshpur Sri Gouranga Vidyapith (H.S); P.O.-Kulitapara, Howrah 711312,West Bengal, India.\\
$^{3}$Department of Physics, Kalyani University, Kalyani, Nadia 741235, West Bengal, India. }


\begin{abstract}
 We study the stability of attractive atomic Bose-Einstein condensate and the macroscopic quantum many-body tunneling (MQT)
in the anharmonic trap. We utilize correlated two-body basis function which keeps all possible two-body correlations. The anharmonic parameter ($\lambda$) is slowly tuned from harmonic to anharmonic. For each choice of $\lambda$ the 
many-body equation is solved adiabatically. The use of the van der Waals
interaction gives realistic picture which substantially differs from 
the mean-field results. For weak anharmonicity, we observe that the 
attractive condensate gains stability with larger number of bosons 
compared to that in the pure harmonic trap. The transition from 
resonances to bound states with weak anharmonicity also differs significantly 
from the earlier study of Moiseyev {\it et.al.}[J. Phys. B: At. Mol. 
Opt. Phys. {\bf{37}}, L193 (2004)]. We also study the tunneling
of the metastable condensate very close to the critical number $N_{cr}$ 
of collapse and observe that near collapse the MQT is the dominant decay mechanism compared to the two-body and three-body loss rate. We also observe the power law behavior in MQT near the critical point. The results for pure harmonic trap are in agreement with mean-field results. However we fail to retrieve the power law behavior in anharmonic trap although MQT is still the dominant decay mechanism.
\end{abstract}
\maketitle
\section{Introduction}
The decay and tunneling of the metastable states is an old quantum 
mechanical problem. However in the context of Bose-Einstein condensate
(BEC)~\cite{Pethick, Wieman1}, it revives new interest 
in the study of decay and tunneling of interacting trapped bosons 
through a potential barrier. Optical traps of finite width can 
support both the bound and resonance states~\cite{Moiseyev,Adhikari,Moiseyev1,Rapedius,Korsch,Carr} 
and the macroscopic quantum tunneling of a BEC through such finite size
barrier is directly experimentally observed~\cite{Shin,Ketterle}. 
Theoretically the  treatment of transport within the mean-field has 
been attempted by many groups~\cite{Moiseyev,Adhikari,Moiseyev1,Rapedius,Korsch,Carr,Paul}. 
Many interesting phenomena are observed due to the nonlinear 
interaction term in the Gross-Pitaeveskii (GP) equation. 
Moiseyev {\it et.al.}~\cite{Moiseyev} have also studied the transition 
from resonance to bound states of trapped attractive BEC when the 
interatomic attraction is increased. Carr {\it et.al.}~\cite{Carr} have studied the 
macroscopic quantum tunneling in a finite potential well in one, two and 
three dimensions. Time-dependent GP equation has been used to study 
the decay process of the BEC~\cite{Adhikari,Paul}. 
In another attempt by the 
group of Heidelberg~\cite{Cederbaum1,Cederbaum2}, the decay and tunneling dynamics of few 
interacting bosons through one-dimensional barrier is studied from the 
first principle and compared with the mean-field results. However all these studies considered only local interaction. But recently it has been reported that for alkali atoms having negative scattering length, one can not neglect the momentum dependence of the scattering cross-section even at very low energy~\cite{Gribakin,Cote}. This implies that one should consider the nonlocal effective potential for attractive BEC. Also near the criticality, the condensate becomes highly correlated and the interatomic correlation can not be neglected. Thus the quantum many-body treatment incorporating the interatomic correlation and a realistic interatomic interaction, which accomodates both the local and nonlocal part of the potential, is necessary. 

In the present work we employ a correlated many-body approach incorporating a realistic interatomic interaction {\it viz.} van der Waals interaction to study the decay and tunneling of the attractive BEC trapped in a quadratic plus quartic confinement. The quartic term takes care of the shallow gaussian potential of finite width.
Such attractive Bose-Einstein condensates are created in routine experiments. The height of the external confining potential well is reduced in a controlled fashion and exponentially screened potential wells are created in the laboratory.  For theoretical calculations the external potential is modeled as $V(r)= \frac{1}{2}m \omega^2 r^2 - \lambda r^4$. In the experiment, quartic confinement is created with blue-detuned  gaussian laser directed along the axial direction and the strength of quartic confinement was $\lambda \sim 10^{-3}$~\cite{Bretin}. Thus for our theoretical study we choose $\lambda$ as a controllable parameter and $\lambda \ll 1.0$. 

In our many-body theory, the choice of van der Waals interaction with a short-range hard core and a $-\frac{C_6}{r^6}$ tail gives the realistic picture. In the many-body effective potential, the metastable BEC is now bounded by 
double barriers of different height in two sides. On the left side of the left barrier there is a narrow deep negative well which is the effect of nonlocality. Thus the BEC may suffer tunneling through both the barriers simultaneously. 
This is distinctly different 
from the tunneling in the mean-field approach where the BEC suffers 
tunneling through the right side barrier which is the effect of finite 
trap size only. In our many-body picture, while the atoms may tunnel out of the trap through the right side barrier, tunneling through the left side barrier corresponds to the collapse. However in our many-body picture the metastable condensate does not collapse truely. Here, due to the presence of deep narrow attractive well on the left of the left side barrier, after tunneling the atoms are accumulated in the deep well and form cluster. In the deep well, density becomes quite large and the rate of two-body and three-body collision is greatly enhanced. Therefore the atoms may acquire enough energy and can be realesed from the trap. On the contrary, it is shown by using the local potential in the mean-field calculation that beyond the critical number, the condensate energy suddenly goes to $-\infty$ and the radius becomes zero. Thus true collapse occurs in this case.
In earlier mean-field calculation MQT from metastable state to the collapsed state is calculated in pure harmonic trap using a trial gaussian wave function and it is shown that near the critical point the tunneling exponent vanishes according to $(1-\frac{N}{N_{cr}})^{5/4}$~\cite{Ueda}, where $N_{cr}$ is the critical number. Now it is obvious to recalculate tunneling exponent using quantum many-body calculation and to extend it for quadratic plus quartic potential. For comparison of MQT rate with other decay mechanism, we also calculate the two-body and three-body loss rate of the BEC in harmonic trap. Our theoretical results show that macroscopic quantum tunneling (MQT) is a dominant decay mechanism near the collapse and we observe much faster increase in the exponential factor in the shallow trap.  In the present work we continuously tune $\lambda$  and study the tunneling process adiabatically. This needs to control the height of the confining potential well in a controlled fashion. As the trapping potential is imposed optically by laser in the routine experiments of BEC, the height of the external potential is controlled by reducing the laser intensity as stated earlier. Thus our present study is quite important as it can be obtained in the laboratory with present day set-up.

The paper is organized as follows. In Section II, we introduce the 
many-body calculation with the correlated harmonic potential basis. 
Section III discuses the numerical results and Section IV concludes 
the summary of our work.

\section{Methodology}
\subsection{Many-body calculation with correlated potential 
harmonic basis}

The earlier theoretical studies on attractive BEC in the harmonic trap 
used the mean-field approximation which results the Gross-Pitaeveskii 
(GP) equation for contact $\delta$-interaction~\cite{Dalfovo}. As the total 
condensate wave function is taken as the product of single particle 
wave functions, the effect of interatomic correlation is completely 
ignored. However specially for the attractive condensate, as the 
atoms come closer and closer, the central density becomes high and 
the condensate becomes highly correlated near the critical point. 
Naturally the interatomic correlation can no longer be ignored 
and one needs a full quantum many-body calculation which takes care 
of the effect of interatomic correlation.

In our present study we solve the many-body Schr\"odinger equation 
by potential harmonic expansion method (PHEM), which basically uses 
a truncated two-body basis set which keeps all possible two-body 
correlation~\cite{Fabre} and we go beyond the mean-field approximation. The 
potential harmonic expansion method with an additional short range 
correlation function, called CPHEM, has already been established as 
a very useful technique for the study of attractive BEC~\cite{Kundu,Anindya1,Anindya2,Sudip}. Here 
we describe the methodology briefly for the interested readers. 
Details are found in our earlier work~\cite{Tapan, Das}.
The Hamiltonian for a system of $A=(N+1)$ identical bosons (each of 
mass $m$) interacting via two-body potential 
$V(\vec{r}_{ij}) = V(\vec{r}_{i}-\vec{r}_{j})$ 
and confined in an external trap (which is modeled as a harmonic 
potential with a quartic term) has the form
\begin{equation}
H=-\frac{\hbar^2}{2m}\sum_{i=1}^{A} \nabla_{i}^{2} 
+ \sum_{i=1}^{A} V_{trap}(\vec{r}_{i}) 
+\displaystyle{\sum_{i,j>i}^{A}} V(\vec{r}_{i}-\vec{r}_{j})\cdot
\end{equation}
After elimination of the center of mass motion and using standard 
Jacobi coordinates~\cite{Fabre,Ballot,MFabre}, the Hamiltonian describing the relative 
motion of the atoms is given by
\begin{equation}
H=-\frac{\hbar^{2}}{m}\sum_{i=1}^{N} 
\nabla_{\zeta_{i}}^{2}+V_{trap} + V_{int}
(\vec{\zeta}_{1}, ..., \vec{\zeta}_{N})\hspace*{.1cm}, 
\end{equation}
$V_{int}$ is the sum of all pair-wise interactions expressed in terms 
of the Jacobi vectors. It is to be noted that Hyperspherical harmonic 
expansion method (HHEM) is an {\it ab-initio} tool to solve the 
many-body Schr\"odinger equation where the total wave function is 
expanded in the complete set of hyperspherical basis~\cite{Ballot}. Although 
HHEM is a complete many-body approach which includes all correlations,
due to large degeneracy of the HH basis, HHEM can not be applied to a 
typical BEC which contains few thousands to few millions of atoms. 
However in the context of experimentally achieved BEC, as the 
interparticle separation is very large compared to the range of 
interatomic interaction, we can safely ignore the effect of three-body and 
higher-body correlation and can keep only the two-body  correlation.
This is perfectly justified for dilute BEC where the probabilities of 
three and higher body collision is negligible. It permits us to 
decompose the total wave function $\Psi$ into two-body Faddeev component 
for the interacting $(ij)$ pair as 
\begin{equation}
\Psi=\sum_{i,j>i}^{A}\phi_{ij}(\vec{r}_{ij},r)\hspace*{.1cm}\cdot
\end{equation}
It is worth to note that $\phi_{ij}$ is a function of two-body 
separation ($\vec{r}_{ij}$) only and also includes the global 
hyperradius $r$, which is given by
$r = \sqrt{\sum_{i=1}^{N}\zeta_{i}^{2}}$.
Thus the effect of two-body correlation comes 
through the two-body interaction in the expansion basis. $\phi_{ij}$ 
is symmetric under $P_{ij}$ for bosonic atoms and satisfy the 
Faddeev equation
\begin{equation}
\left[T+V_{trap}-E_R\right]\phi_{ij}
=-V(\vec{r}_{ij})\sum_{kl>k}^{A}\phi_{kl}
\end{equation}
where $T$ is the total kinetic energy. Operating $\sum_{i,j>i}$ on both 
sides of equation (4), we get back the original Schr\"odinger equation. 
In this approach, we assume that when ($ij$) pair interacts, the rest 
of the bosons are inert spectators. Thus the total hyperangular momentum 
quantum number as also the orbital angular momentum of the whole system 
is contributed by the interacting pair only. Next  the  $(ij)$th Faddeev 
component is expanded in the set of potential harmonics (PH) (which is 
a subset of HH basis and sufficient for the expansion of $V(\vec {r}_{ij})$) 
appropriate for the ($ij$) partition as 
\begin{equation}
\phi_{ij}(\vec{r}_{ij},r)
=r^{-(\frac{3N-1}{2})}\sum_{K}{\mathcal P}_{2K+l}^{lm}
(\Omega_{N}^{ij})u_{K}^{l}(r) \hspace*{.1cm}\cdot
\end{equation}
$\Omega_N^{ij}$ denotes the full set of hyperangles in the $3N$-dimensional 
space corresponding to the $(ij)$th  interacting pair and 
${\mathcal P}_{2K+l}^{lm}(\Omega_N^{ij})$ is called the PH basis. It 
has an analytic expression:
\begin{equation}
{\mathcal P}_{2K+l}^{l,m} (\Omega_{N}^{(ij)}) =
Y_{lm}(\omega_{ij})\hspace*{.1cm} 
^{(N)}P_{2K+l}^{l,0}(\phi) {\mathcal Y}_{0}(D-3) ;\hspace*{.5cm}D=3N ,
\end{equation}
${\mathcal Y}_{0}(D-3)$ is the HH of order zero in 
the $(3N-3)$ dimensional space spanned by $\{\vec{\zeta}_{1}, ...,
\vec{\zeta}_{N-1}\}$ Jacobi vectors; $\phi$ is the hyperangle given by
$r_{ij}$ = $r\hspace*{0.1cm} cos\phi$. For the remaining $(N-1)$
noninteracting bosons we define hyperradius as
\begin{eqnarray}
 \rho_{ij}& = &\sqrt{\sum_{K=1}^{N-1}\zeta_{K}^{2}}\nonumber\\
          &= &r \sin\phi \hspace*{.01 cm}\cdot
\end{eqnarray}
such that $r^2=r_{ij}^2+\rho_{ij}^2$ and $r$ represents the global 
hyperradius of the condensate. The set of $(3N-1)$ quantum 
numbers of HH is now reduced to {\it only} $3$ as for the $(N-1)$ 
non-interacting pair
\begin{eqnarray}
l_{1} = l_{2} = ...=l_{N-1}=0,   & \\
m_{1} = m_{2}=...=m_{N-1}=0,  &   \\
n_{2} = n_{3}=...n_{N-1} = 0, & 
\end{eqnarray}
and for the interacting pair $l_{N} = l$, $m_{N} = m$ and  $n_{N} = K$.
Thus the $3N$ dimensional Schr\"odinger equation reduces effectively
to a four dimensional equation with the relevant set of quantum 
numbers: hyperradius $r$, orbital angular momentum quantum number $l$,
azimuthal quantum number $m$ and grand orbital quantum number $2K+l$
 for any $N$.
Substituting in Eq(4) and projecting on a particular PH, a set of 
coupled differential equation (CDE) for the partial wave $u_{K}^{l}(r)$
is obtained
\begin{equation}
\begin{array}{cl}
&\Big[-\frac{\hbar^{2}}{m} \frac{d^{2}}{dr^{2}} +
V_{trap}(r) + \frac{\hbar^{2}}{mr^{2}}
\{ {\cal L}({\cal L}+1) \\
&+ 4K(K+\alpha+\beta+1)\}-E_R\Big]U_{Kl}(r)\\
+&\displaystyle{\sum_{K^{\prime}}}f_{Kl}V_{KK^{\prime}}(r)
f_{K^{\prime}l}
U_{K^{\prime}l}(r) = 0
\hspace*{.1cm},
\end{array}
\end{equation}\\
where ${\mathcal L}=l+\frac{3A-6}{2}$, $U_{Kl}=f_{Kl}u_{K}^{l}(r)$, 
$\alpha=\frac{3A-8}{2}$ and $\beta=l+1/2$.\\
$f_{Kl}$ is a constant and represents the overlap of the PH for
interacting partition with the sum of PHs corresponding  to all 
partitions~\cite{MFabre}.
The potential matrix element $V_{KK^{\prime}}(r)$ is given by
\begin{equation}
V_{KK^{\prime}}(r) =  
\int P_{2K+l}^{lm^*}(\Omega_{N}^{ij}) 
V\left(r_{ij}\right)
P_{2K^{\prime}+1}^{lm}(\Omega_{N}^{ij}) d\Omega_{N}^{ij} 
\hspace*{.1cm}\cdot
\end{equation}

\subsection{Introduction of additional short range correlation}

As pointed earlier in the mean-field GP equation the two-body interaction is represented 
by a single parameter, the $s$-wave scattering length $a_{sc}$ only. 
It disregards the detailed structure. The presence of essential 
singularity as $r \rightarrow 0$  for the attractive contact $\delta$-interaction 
makes the Hamiltonian unbound from below. So in our present many-body 
calculation, we use a realistic interatomic potential like the 
van der Waals potential with an attractive $-\frac{1}{r^6}$ tail at 
large separation and a strong short range repulsion. The inclusion of 
detailed structure in the two-body potential with the 
short range repulsive core needs to include 
an additional short range correlation in the PH basis. This short 
range behavior is represented by a hard core of radius $r_c$ and we 
calculate the two-body wave function $\eta(r_{ij})$ by solving the 
zero-energy two-body Schrodinger equation
\begin{equation}
-\frac{\hbar^2}{m}\frac{1}{r_{ij}^2}\frac{d}{dr_{ij}}\left(r_{ij}^2
\frac{d\eta(r_{ij})}{dr_{ij}}\right)+V(r_{ij})\eta(r_{ij})=0
\hspace*{.1cm}\cdot
\end{equation} 
This zero-energy two-body wave function $\eta(r_{ij})$ is a good 
representation of the short range behavior of $\phi_{ij}$ as in the 
experimental BEC, the energy of the interacting pair is negligible 
compared with the depth of the interatomic potential. It is 
taken as the two-body correlation function in the PH expansion 
basis. The value of $r_c$ is obtained from the requirement that 
the calculated $a_{sc}$ has the expected value~\cite{Kundu}. We 
introduce this as a short-range correlation function in the expansion 
basis. This also improves largely the rate of convergence of the 
PH basis and we call it as correlated Potential Harmonic expansion 
method (CPHEM). We replace Eq(5) by 
\begin{equation}
\phi_{ij}(\vec{r}_{ij},r)
=r^{-(\frac{3N-1}{2})}\sum_{K}{\mathcal P}_{2K+l}^{lm}
(\Omega_{N}^{ij})u_{K}^{l}(r) \eta(r_{ij}) \hspace*{.1cm}\cdot
\end{equation}
and the correlated PH (CPH) basis is given by
\begin{equation}
[{\mathcal P}_{2K+l}^{l,m} (\Omega_{N}^{(ij)})]_{correlated} =
 {\mathcal P}_{2K+l}^{l,m} (\Omega_{N}^{(ij)}) \eta(r_{ij}) ,
\end{equation}
The correlated potential matrix $V_{KK^{\prime}}(r)$ is now given by
\begin{equation}
\begin{array}{cl}
&V_{KK^{\prime}}(r) =(h_{K}^{\alpha\beta} h_{K^{\prime}}^
{\alpha\beta})^{-\frac{1}{2}}\times \\
&\int_{-1}^{+1} \{P_{K}^{\alpha\beta}(z) 
V\left(r\sqrt{\frac{1+z}{2}}\right)
P_{K^{\prime}}^{\alpha \beta}(z)\eta\left(r\sqrt{\frac{1+z}{2}}\right)
W_{l}(z)\} dz \hspace*{.1cm}\cdot
\end{array}
\end{equation}
Here $P_{K}^{\alpha\beta}(z)$ is the Jacobi polynomial, and its 
norm and 
weight function are $h_{K}^{\alpha\beta}$ and $W_{l}(z)$   
respectively~\cite{Abramowitz}.

One may note that the inclusion of $\eta(r_{ij})$ makes the PH basis 
non-orthogonal. One may surely use the standard procedure for handling 
non-orthogonal basis. However in the present calculation we have 
checked that $\eta(r_{ij})$ differs from a constant value only by small 
amount and the overlap $\Big< {\mathcal P}_{2K+l}^{l,m} (\Omega_{N}^{(ij)})|{\mathcal P}_{2K+l}^{l,m} (\Omega_{N}^{(kl)})\eta(r_{kl})\Big>$ is quite small. Thus we get back the Eq(11) 
approximately when the correlated potential matrix is calculated by Eq(16). 

This is important to note that our correlated basis function keeps all possible two-body correlation only. The natural question is also the significant role played by the three-body correlation. However in our present study we keep the scattering length $a_{sc} \sim 10^{-4}$ o.u. ($\approx 10$ Bohr) which means that although the condensate is correlated near the critical point of collapse but it is still extremely dilute, $na_{sc}^{3}\ll 1$ (where $n$ is the density). Thus the interatomic correlation in the two-body level is significant which is also observed in our earlier study~\cite{Anindya2}. However near the Feshbach resonance where the scattering length is order of few thousands Bohr the three and higher-body correlation must be taken into account.
 
\section{Results}

\subsection{Choice of two-body potential and calculation of 
many-body effective potential}

In the standard treatment of alkali-metal atoms, the atom-atom interaction is usually replaced by an effective zero-range pseudo-potential as $V(r)=g \delta^3(r)$, where $g$ is the strength of the contact potential and is given by $g=\frac{4\pi\hbar^2a_{sc}}{m}$, $a_{sc}$ is the $s$-wave scattering length. By using such an effective potential one gets the familiar Gross-Pitaevskii equation. However few alkali atoms are characterised by the negative scattering length at low energy. $^{85}$Rb and $^{7}$Li are two such alkali atoms for which BEC is experimentally observed when the number of atoms is below the critical number. Out of these two $^{85}$Rb is chosen as the best candidate where we can tune the atom-atom inetraction using magnetic field to induce Feshbach resonance. This allows one to study the onset of instability in a controlled way~\cite{Cornish,Roberts}. The possibility of having negative as well as positive scattering length in different atomic systems arises because the interatomic potential may support bound states. Under such condition the scattering length will be momentum dependent~\cite{Gribakin,Cote}. The energy-dependent scattering cross-section significantly deviates from its zero-energy limit even at very low energy ~\cite{Gribakin,Cote}. Thus we can not neglect the momentum dependence of the effective potential for the colliding atoms. This tells us that the effective potential is no more local, it is nonlocal changing from attractive to repulsive at a characteristic range $r_e$. To accomodate this kind of nonlocal interaction in the earlier mean-field calculation, it is assumed that the attractive potential has a finite range $r_e$ and the repulsive part is modelled by using a local positive term~\cite{Luca,Parola}. Thus the effective interaction is $V(r)=\frac{4\pi\hbar^2}{m}[a_r + (a_s-a_r)f(kr_e)]$ where $a_s$ corresponds to the attractive potential and the repulsive part is characterised by $a_r$. The shape function is chosen as either Lorenzian or Gaussian. This gives rise to the nonlocal GP energy function~\cite{Parola}. In our present calculation we model the interatomic interaction by a more realistic long-range potential $-$ the van der Waals potential with a hard core of radius $r_{c}$, viz., 
$V(r_{ij}) = \infty$ for $r_{ij} \le r_{c}$ and $= -\frac{C_6}{r_{ij}^6}$
for $r_{ij} > r_{c}$, which can accomodate both the local and nonlocal part of the potential. $C_6$ is known for a 
specific atom and in the limit of $C_6 \rightarrow 0$, the potential 
becomes a hard sphere and the cutoff radius exactly coincides with the
$s$-wave scattering length $a_{sc}$. In our choice
of two-body potential we tune $r_c$ to reproduce the experimental 
scattering length $a_{sc}$. As we decrease $r_c$, $a_{sc}$ decreases and at a 
particular critical value of $r_c$  it passes through $- \infty$ to $\infty$~\cite{Kundu}.
For our present calculation we choose $^{85}$Rb atoms with $C_6$ = $6.4898 \times 10^{-4}$ o.u.~\cite{Pethick} 
and  tune $r_c$ to obtain $a_{sc}$ = $-1.832 \times 10^{-4}$ o.u., which is one of the choices of scattering length $a_{sc}$ in the controlled collapse experiment of Roberts {\it et.al.}~\cite{Roberts, Cornish}. 
We choose $r_c$ such that it corresponds to the zero node in the two-body 
wave function. The chosen value of $r_c$ for our calculation is 
$1.3955 \times 10^{-3}$ o.u.. With these set of parameters  
we solve the coupled differential equation by
hyperspherical adiabatic approximation (HAA)~\cite{Coelho}. In HAA, 
we assume that the hyperradial motion is slow compared to the 
hyperangular motion and the potential matrix together with the 
hypercentrifugal repulsion is diagonalized for a fixed value of $r$. 
Thus the effective potential for the hyperradial motion is obtained 
as a parametric function of $r$. We choose the lowest eigen potential 
$\omega_0(r)$ as the effective potential in which the condensate moves
collectively. The energy and wave function of the condensate are 
finally obtained by solving the adiabatically separated hyperradial equation 
in the extreme adiabatic approximation (EAA)
\begin{equation}
 \left[-\frac{\hbar^{2}}{m}\frac{d^{2}}{dr^{2}}+\omega_{0}(r)-E_{R}
\right]\zeta_{0}(r)=0\hspace*{.1cm},
\end{equation}
subject to approximate boundary conditions on $\zeta_0(r)$.
For our numerical calculation we fix $l=0$ and truncate the CPH basis 
to a maximum value $K=K_{max}$ requiring proper convergence. 
\begin{figure}
  \begin{center}
    \begin{tabular}{cc}
      \resizebox{80mm}{!}{\includegraphics[angle=0]{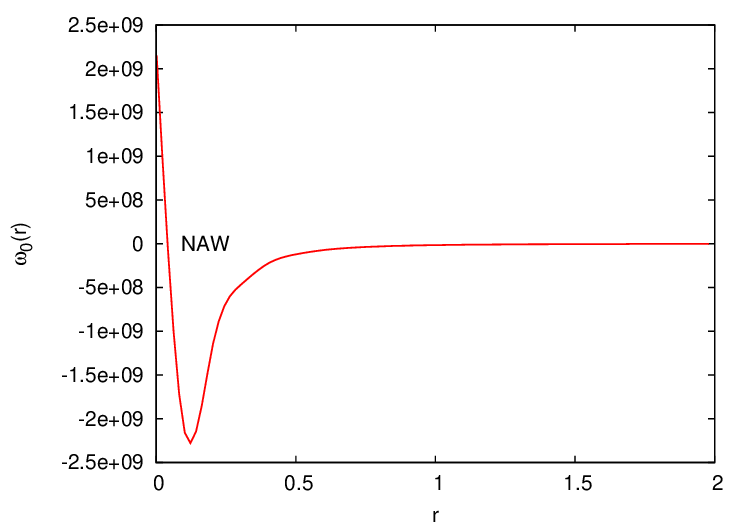}} &\\
         (a)  &\\
      \resizebox{80mm}{!}{\includegraphics[angle=0]{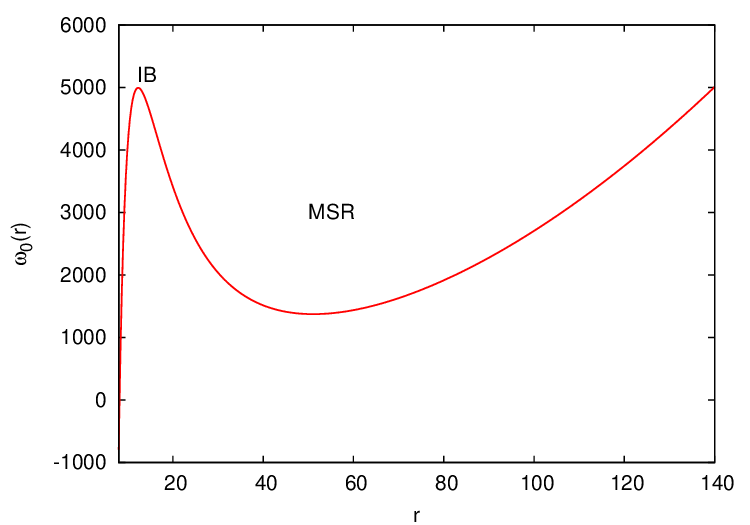}} & \\
         (b)  &\\
    \end{tabular}
  \end{center}
\caption{(Color online) Plot of effective potential $\omega_0(r)$
of an attractive BEC with $N=1000$ in a pure harmonic trap. In panel (a) 
the deep narrow attractive well NAW is ploted while in panel (b) the 
metastable region MSR is presented.}
\end{figure}

The advantages of the CPHEM are as follows.
\begin{enumerate}
\item[i)] As for the attractive BEC, the condensate becomes highly 
correlated, the choice of CPH basis (which keeps all possible 
two-body correlations) is perfectly justified for the description 
of correlated dilute BEC.
\item[ii)] By using the HAA we basically reduced the $3N$ dimensional 
problem into an effective one-dimensional problem in hyperradial space 
which provides both qualitative and quantitative picture of the system. 
As the effective quantum numbers are always {\it four} (for any $N$) 
we can run our many-body code for very high values of $N$.
\item[iii)] Our PHEM method is in no way restricted to any particular choice of two-body interaction potential. And we have already used different two-body potentials, e.g. a Gaussian potential~\cite{Tapan}, van der Waals potential~\cite{Sudip,Anindya1,Anindya2,Pankaj1,Pankaj2}, LM2M2 and TTY potential (for $^4$He trimer)~\cite{Barnali,SudipF} for different systems in different context.
\item[iv)] Due to the use of realistic nonlocal two-body potential the many-body effective potential strongly differs from the mean-field potential. In GP, the choice of contact $\delta$-interaction 
in the two-body potential gives rise to the pathological singularity
in the effective potential. Thus the study of post-collapse scenario of 
the attractive condensate is beyond the scope of mean-field theory as mentioned earlier. 
Whereas the presence of short-range hard core in the van der Waals 
interaction not only removes the singularity, it also gives the 
realistic scenario and can describe the formation of atomic cluster 
after collapse.
\end{enumerate}

\subsection{Macroscopic quantum tunneling (MQT) with tuned $\lambda$}

As stated earlier, our choice $V(r)= \frac{1}{2} m \omega^{2} r^{2} - \lambda r^{4}$ models the optical trap used in many experiments~\cite{Stock,Bretin}. In the routine experiments, after the formation of stable BEC, the height of the potential well is gradually reduced by reducing the laser intensity in the optical trap. Thus the attractive BEC is created in exponentially screened potential well and will exhibit different kinds of tunneling phenomena. Thus to corroborate with the real experimental situation, we take $\lambda$ as a controlable parameter and tune it in a controlled fashion. We may assume that the time dependent $\lambda$ is a constant to leading order as $\lambda(t)=\lambda_0(1+\epsilon_0(t))$, where $\lambda_0$ is chosen as the anharmonic strength parameter $\approx 10^{-6} - 10^{-4} \ll 1.0$ and $\epsilon_0(t)$ is the fluctuation of the laser intensity which is controlled externally. Thus the use of time dependent anharmonic strength factor will need  the solution of full time-dependent many-body Schr\"odinger equation. However for our present study we solve the Schr\"odinger equation adiabatically for each choice of $\lambda$ and observe interesting features in MQT of attractive condensate.
In Fig.~1 we plot the many-body effective potential $\omega_0(r)$ as a function 
of hyperradius $r$ for $N= 1000$ atoms in the pure harmonic trap ($\lambda = 0$). For 
$N < N_{cr} (\approx 2483)$, the condensate is metastable and is associated 
with a deep and narrow attractive well (NAW) on the left side. For 
$r \rightarrow 0$, there is a strong repulsive wall which is the 
reflection of the hard core van der Waals interaction and shows the effect of nonlocality as stated above. For $N$ less than 
the critical number $N_{cr}$ a metastable region (MSR) appears for 
larger $r$. An intermediate barrier (IB) separates the NAW from the MSR. 
In the panel (a) of Fig. 1, the NAW together with the repulsive core is shown. 
The IB and MSR have been shown in panel (b) of Fig. 1. With increase in $N$, 
we observe that the height of IB decreases, together with a decrease 
in the difference ($\Delta \omega$) between the maximum of IB ($\omega_{max}$) 
and minimum of MSR ($\omega_{min}$) and the NAW starts to be more 
negative and narrower.  As $N \rightarrow N_{cr}$, $\omega_{max}$ 
and $\omega_{min}$ coincides, the MSR disappears. In the earlier attempt by Ueda and Leggett~\cite{Ueda}, it has been shown that there is MQT from the metastable state to the collapsed state near the criticality by using only the local effective potential. At $N \approx N_{cr}$, the radius of the condensate becomes zero and the energy becomes $-\infty$ as the Hamiltonian is unbound from below. However using the nonlocal potential we have a finite deep and narrow negative well (NAW) on the left hand side which can accomodate the condensate even when $N > N_{cr}$. As $N$ becomes larger than $N_{cr}$, all the atoms now get trapped in NAW and we have a self-trapped system which gives rise to a cluster state. Thus in the pre-collapse state we have metastable BEC in the positive well (MSR) on the right side which we describe as the low-density branch. In the post-collapse state we have atomic cluster in the NAW and we describe it as high-density branch. To be more quantitative we calculate the average size $r_{av}$ for both branches as~\cite{Anasua}
\begin{equation}
r_{av} = \Big<\frac{1}{A}\sum_{i=1}^{A}(\vec{x_{i}}-
\vec{R})^{2}\Big>^{1/2} = \frac{<r^2>^{1/2}}{\sqrt{2A}},
\end{equation}
where $\vec{R}$ is the center of mass coordinate. In Fig.~2 we plot $r_{av}$ as a function of $N$. The upper branch corresponds to the metastable condensate before collapse which shows a sharp fall in $r_{av}$ near $N=N_{cr}\approx2483$. The lower branch corresponds to the high-density stable branch in the NAW which starts from $N=50$. The size of the many-body state in NAW is $\sim 0.0027 \mu$m which is of the order of the size of atomic cluster. Thus the use of the realistic van der Waals interaction not only shows the nonlocal effect but also we get some additional new features compared to the local potential. In case of local interaction, the MQT occurs from the metastable state to the  collapsed state whereas in our realistic calculation MQT occurs from the low-density branch to the high-density branch. This is to point out  that the high-density branch is mechanically stable and has very short life-time. Due to two-body dipolar collision and three-body recombinations, the atoms may acquire significant energy and can be released from the trap.
\begin{figure}[hbpt]
\vspace{-10pt}
\centerline{
\hspace{-3.3mm}
\rotatebox{0}{\epsfxsize=8.8cm\epsfbox{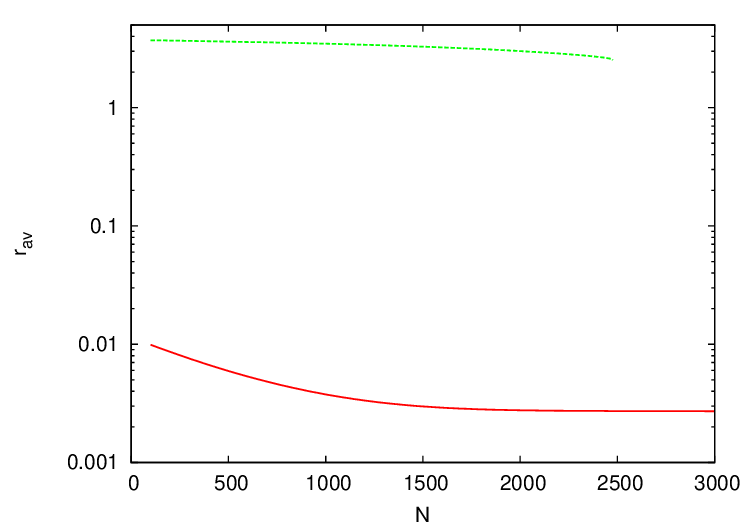}}}
\caption{(color online)Plot of $r_{av}$ (in $\mu$m) as a function of number of bosons $N$ in harmonic trap. The upper green dashed curve corresponds to the low-density branch in MSR and the lower red smooth curve corresponds to the high-density branch in NAW.}
\end{figure}

This is also to be noted that the observed high-density branch can be correlated with the fragmented metastable state in the multi-orbital mean-field theory~\cite{Cederbaum1,Cederbaum2,Cederbaum}. The usual GP theory is a single orbital mean-field theory whereas in the multi-orbital mean-field theory the bosons are distributed among several orthogonal states. It has been shown that the attractive BEC may possess metastable states for $N$ much larger than the critical number obtained in GP theory~\cite{Dalfovo}. This is similar to the existance of high-density branch in the narrow attractive well (NAW) in our calculation.    . 

The important issue is the study of MQT near the critical point and to observe the power law behavior in MQT rate. We calculate the WKB tunneling rate as 
\begin{eqnarray}
\Gamma_{N}^{tunnel}& = &N\nu\exp(-2\int_{r_1}^{r_2}
\sqrt{2[\omega_0(r)-E]}\hspace*{.1cm}dr)\nonumber \\
                   & = & N\nu T \hspace*{.1cm}\cdot
\end{eqnarray}
where the limits of the integration $r_1$ and $r_2$ are the inner and outer turning points of the metastable region of $\omega_0(r)$. $T$ is the WKB tunneling probability and $\nu$ is the frequency of impact. Though WKB is a semiclassical approximation method, its use in our present study can be justified as follows. In BEC, below the critical temperature $T_c$ there is a macroscopic occupation of bosons in the lowest quantum state, the de-Broglie wave lengths of the neighbouring atoms overlap and the individual atoms loose their quantum identity. Thus the whole condensate is treated as a single quantum stuff which is described by the condensate wave function $\Psi$. In our many-body picture, the condensate moves in the effective potential $\omega_0(r)$ in the hyperradial space as a single quantum entity. Thus we look for the macroscopic behavior of the collective motion of the condensate. Thus effect of quantum fluctuations does not appear in our methodology. The thermal fluctuation is also absent as we consider only zero-temperature BEC. This also justifies the use of WKB approximation. The similar kind of attempt has been taken in the earlier calculation of Bohn {\it et al}~\cite{Bohn}. Comparing with the notations of Ref.~\cite{Ueda} we may express MQT as $\Gamma_{N}^{tunnel}=A\exp(-\frac{S^{B}}{\hbar})$ with $A\equiv N\nu$ and $\frac{S^{B}}{\hbar} \equiv 2\int_{r_1}^{r_2}\sqrt{2[\omega_0(r)-E]}\hspace*{.1cm}dr$. In Fig.~3 we plot $\ln (\frac{S^{B}}{\hbar})$ as a function of $\ln (1-\frac{N}{N_{cr}})$ near the critical point and we obtain
\begin{equation}
\frac{S^{B}}{\hbar} = 5.46 N (1-\frac{N}{N_{cr}})^{1.26}
\end{equation} 
\begin{figure}[hbpt]
\vspace{-10pt}
\centerline{
\hspace{-3.3mm}
\rotatebox{0}{\epsfxsize=8.8cm\epsfbox{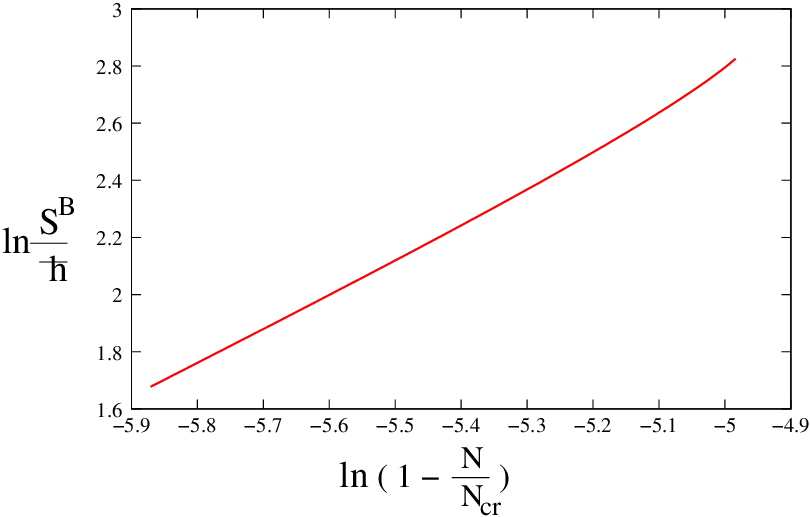}}}
\caption{(color online)Plot of $\ln (\frac{S^{B}}{\hbar})$ as a function of $\ln (1-\frac{N}{N_{cr}})$ for $\lambda=0$.} 
\end{figure}

Comparing Eq.~(20) with Eq.~(25) of Ref.~\cite{Ueda} we determine the numerical coefficient as $5.46$ and the tunneling exponent near the critical point is $1.26$. Thus the power law behavior of MQT is in close agreement with earlier mean-field results~\cite{Ueda} and it again confirms that MQT is a dominant decay mechanism of the attractive condensate near the critical point. Next to quantify the MQT, we present some specific values in Table I. We observe that with $N \approx N_{cr}$, $1-\frac{N}{N_{cr}} \sim 10^{-3}$ and the MQT rate is significant. Whereas for $N < N_{cr}$,  MQT is negligible. The important question in this direction is to compare the decay rate of the condensate due to MQT with the loss rate $\Gamma_N^{collision}$ due to the two-body and three-body collisions. We calculate the two-body dipolar loss rate and three-body recombination by~\cite{Pethick} 
\begin{equation}
 \Gamma_N^{collision} = K_2 \int d\tau |\Psi|^{4} + K_3 \int d\tau |\Psi|^{6}\hspace*{.1cm}
\end{equation}
Here $K_2$ is the two-body dipolar loss rate coefficient and has the value $(1.87 \pm 0.95 \pm 0.19) \times 10^{-14}$ cm$^3$/sec. The three-body recombination loss rate coefficient $K_3 = (4.24^{+0.70}_{-0.29} \pm 0.85) \times 10^{-24}$ cm$^6$/sec~\cite{Wieman}.
We present the values of $\Gamma_N^{tunnel}$ and $\Gamma_N^{collision}$ for various $N$ close to $N_{cr}$ in the Table I. We find that near the critical point $N_{cr}$ the MQT rate $\Gamma_N^{tunnel}$ increases much faster comapared to the loss rate $\Gamma_N^{collision}$ due to the two-body and three-body collisions and just before collapse $\Gamma_N^{tunnel}$ is $10^{4}$ times larger than $\Gamma_N^{collision}$. Thus MQT is the most significant decay mechanism for the attractive condensate near the critical point. However away from the critical point $\Gamma_N^{tunnel}$ and $\Gamma_N^{collision}$ are comparable and far away from the criticality, the condensate in the MSR is highly stable showing almost vanishing value of $\Gamma_N^{tunnel}$.  

Next we slowly change $\lambda$ from very close to harmonic to very small anharmonic ($\lambda \sim 10^{-6}$). For such weak anharmonicity, the metastable condensate in the MSR is tightly bounded by the high intermidiate barrier as before in pure harmonic trap. However the intermediate barrier on the left side gradually increases with very small incease in $\lambda$ which indicates the greater stability of the condensate. In pure isotropic and harmonic trap $(\lambda=0)$, we have observed that collapse occurs at $N_{cr}=2483$, the MSR disappears when $N = N_{cr}$. However with small anharmonicity $\sim 10^{-6}$, the MSR starts to develop and the metastable condensate reappears. 
To describe the condensate stability we calculate the condensate radius $r_{av}$ by Eq.~(18) and plot it as a function of $\lambda$($\sim 10^{-6}$) with fixed $N$ in Fig.~4. The sharp increase in $r_{av}$ signifies the greater stability of the condensate. In this connection we calculate the stability factor defined as $k_{cr}=\frac{N_{cr}|a_{s}|}{a_{ho}}$. We use it as in our many-body picture we consider the collective behaviour of the condensate and identity of the individual atoms is completely lost after the formation of the effective potential. This is the usual picture of hyperspherical adiabatic approximation (HAA) which is frequently applied in different atomic and nuclear problems. Thus the effect of interparticle correlation is taken into account to obtain the many-body effective potential where the whole condensate is treated as single entity. Also we consider only the zero-temperature BEC. Therefore both the quantum fluctuation and thermal fluctuation are absent as stated earlier. This implies that the depletion is negligibly small for the zero-temperature BEC in our many-body picture. The effective potential is further used to calculate the $N_{cr}$ where the low-density metastable branch makes transition to the high-density branch. This value of $N_{cr}$ is directly used in the above expression of $k_{cr}$. For $\lambda=0$, in pure harmonic trap, $k_{cr}=0.456$ which is in very close agreement with the experimental value~\cite{Roberts} and gives better result than the mean-field theory. In Fig.~5 we plot $k_{cr}$ against $\lambda$ and the smooth increase in the stability factor with $\lambda$ signifies the greater stability of the condensate in the anharmonic trap.
\begin{figure}[hbpt]
\vspace{-10pt}
\centerline{
\hspace{-3.3mm}
\rotatebox{0}{\epsfxsize=8.8cm\epsfbox{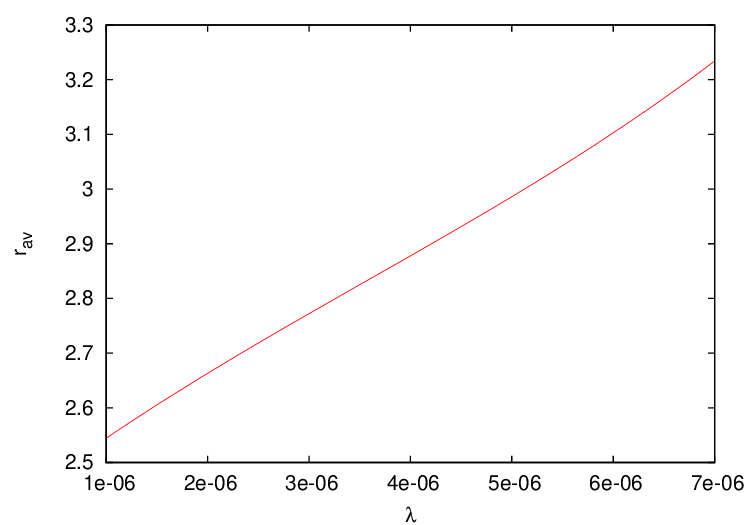}}}
\caption{(color online) Plot of average size of the condensate $r_{av}$ (in $\mu$m) against $\lambda$ (in $o.u.$) for a $N=2483$.}
\end{figure}
\begin{figure}[hbpt]
\vspace{-10pt}
\centerline{
\hspace{-3.3mm}
\rotatebox{0}{\epsfxsize=8.8cm\epsfbox{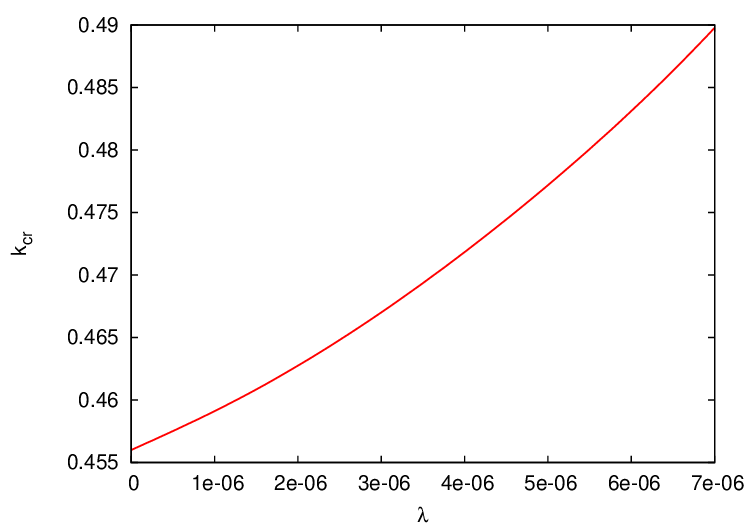}}}
\caption{(color online) Plot of the stability factor $k_{cr}$ against $\lambda$.}
\end{figure}
\begin{center}
\begin{table}[!]
\caption{The values of tunneling exponent, MQT rate $\Gamma_{N}^{tunnel}$ and  two-body and three-body loss rate $\Gamma_N^{collision}$ for various $N$ near $N_{cr}$ for pure harmonic trap ($\lambda=0$).}
\begin{tabular}{|l|l|l|l|}
 \hline
N &  $\frac{S^{B}}{\hbar}$ & $\Gamma_{N}^{tunnel}$ & $\Gamma_N^{collision}$\\ \hline
2476 & 5.35   & 1.02  & $1.682 \times 10^{-4}$\\ \hline
2474 & 5.799  & 0.673 &	$1.595 \times 10^{-4}$\\ \hline
2472 & 11.65  &  $2.041 \times 10^{-3}$ & $1.576 \times 10^{-4}$\\ \hline
2468 & 13.17  &  $4.818 \times 10^{-4}$ & $1.49 \times 10^{-4}$\\ \hline
2466 & 16.869 &  $1.341 \times 10^{-5}$ & $1.475 \times 10^{-4}$\\\hline
2400 & 138.07 &  $3.891 \times 10^{-58}$& $1.203 \times 10^{-4}$\\ \hline
\end{tabular}
\end{table}
\end{center}

In the earlier study of Moiseyev {\it et.al.}~\cite{Moiseyev} 
the transition from resonance to bound state has been studied when the 
attractive interaction gradually increases. However to study such 
transition, an additional negative offset ($V_0$) was required 
to prevent collapse. For $V_0=0$, the authors found  in the Ref.~\cite{Moiseyev} 
that resonance never turns into bound state as the nonlinearity becomes 
more negative, the condensate collapses. However due to the use of hard core 
van der Waals potential having long range attractive tail we do not require any such external offset to 
create the metastable condensate. Thus the increase in stability and the 
transition from resonance to quasi-bound states for very 
weak anharmonicity strongly differs from the earlier observation of 
Moiseyev {\it et.al.}.
Further to observe the power law behavior in quantum tunneling, we plot $\ln (\frac{S^{B}}{\hbar})$ as a function of $\ln (1-\frac{N}{N_{cr}})$ for small anharmonicity in Fig.~6. We fail to retrieve the power law behavior for $\lambda \neq 0$.
\begin{figure}[hbpt]
\vspace{-10pt}
\centerline{
\hspace{-3.3mm}
\rotatebox{0}{\epsfxsize=8.8cm\epsfbox{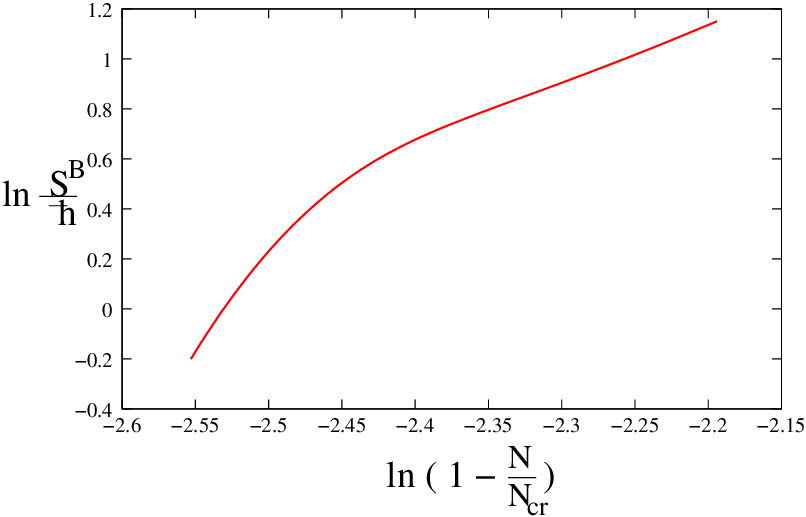}}}
\caption{(color online) Plot of $\ln (\frac{S^{B}}{\hbar})$ as a function of $\ln (1-\frac{N}{N_{cr}})$ for $\lambda \sim 10^{-6}$.}
\end{figure}
\begin{center}

\begin{table}[!]
 \caption{MQT rate through both the barriers for various $N$ near the criticality for $\lambda \sim 10^{-5}$.}
\begin{tabular}{|l|l|l|l|l|}
\hline
N & \multicolumn {2}{|c|}{Left Tunneling}
& \multicolumn{2}{|r|}{Right Tunneling}\\ \cline{2-5}
  & $\frac{S^B}{\hbar}$ & $\Gamma^{tunnel}_{N}$ & $\frac{S^B}{\hbar}$ & $\Gamma^{tunnel}_{N}$\\ \hline
2772 & 6.87 & 7.34 & 18.06 & 1.28 $\times 10^{-5}$\\ \hline
2771 & 8.58 & 1.35 & 18.89 & 2.15 $\times 10^{-5}$\\ \hline
2770 & 10.18 & 0.28 & 19.63 & 4.60 $\times 10^{-5}$\\ \hline
2769 & 11.92 & 4.97 $\times 10^{-2}$ & 20.13 & 1.06 $\times 10^{-5}$\\ \hline
\end{tabular}
\end{table}
\end{center}

With further increase in $\lambda$ ($\sim 10^{-5}$) the condensate in the MSR is now bounded by two barriers of finite width and may suffer tunneling from the MSR through both the barriers simultaneously. In Fig.~7 we plot the many-body effective potential showing two intermediate barriers. We have observed two criticalities associated with these two barriers for a fixed $\lambda$ but varying $N$ $-$ the first criticality corresponds to the right barrier and the second criticality corresponds to the left barrier~\cite{Sudip}. We present the MQT through both the barriers in Table II. We observe that near the critical point when $1-\frac{N}{N_{cr}} \sim 10^{-3}$, the MQT through the left barrier increases enormously with increase in particle number, whereas for the same set of particle numbers, MQT through right side barrier is almost insignificant. Thus although our theoretical many-body calculation presents the possibility of dual tunneling through two adjacent barriers, however from the experimental point of view, MQT is significant through left side only when $\lambda\sim 10^{-5}$. We also observe a close interplay between the number of particles in the trap and the anharmonic interaction and it reflects significant change in barrier heights. If $\lambda$ is increased further slightly (but still $\sim 10^{-5}$), keeping the particle number fixed, we observe that the height of the right barrier decreases and that of the left barrier increases. This enhances the MQT through the right barrier significantly and the MQT through the left barrier decreases. Thus from our present study one may conclude that the MQT becomes a significant decay mechanism for the anharmonicity strength $\lambda \sim 10^{-5}$ as tunneling through either left side or right side barrier would be measured experimentally. However our many-body calculation exhibits double branches if we plot WKB tunneling probability $T$ as a function of $\lambda$ (see Fig.~ 8). We observe a critical point of $\lambda$ below which $T_{left}$ (tunneling through left barrier) is significant whereas  above which $T_{right}$ (tunneling through right side barrier) becomes significant. At exactly $\lambda$=$\lambda_{critical} \approx 8.99 \times 10^{-6}$, the two branches cross each other and $T_{left}=T_{right}$. From the view point of the semiclassical calculation we may point out that at $\lambda$=$\lambda_{critical}$, the two barriers on left and right side will be of same height. Although in present day experiments, only the significant MQT rate is determined, however our findings exhibit possibility of dual tunneling. In near future, using highly controlled laser beam it may be possible to probe the critical strength of quartic confinement as found in our calculation.
\begin{figure}[hbpt]
\vspace{-10pt}
\centerline{
\hspace{-3.3mm}
\rotatebox{0}{\epsfxsize=8.8cm\epsfbox{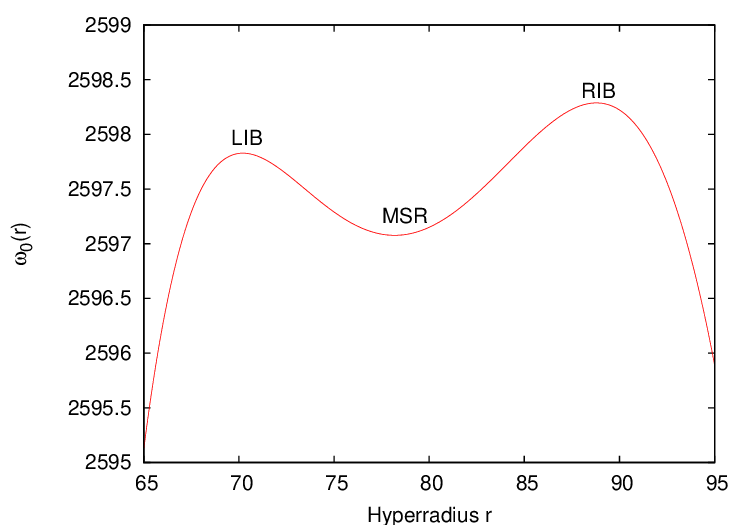}}}
\caption{(color online) Plot of the effective potential $\omega_0(r)$ for $\lambda \sim 10^{-5}$ and $N=2760$.}
\end{figure}

 \begin{figure}[hbpt]
\vspace{-10pt}
\centerline{
\hspace{-3.3mm}
\rotatebox{0}{\epsfxsize=8.8cm\epsfbox{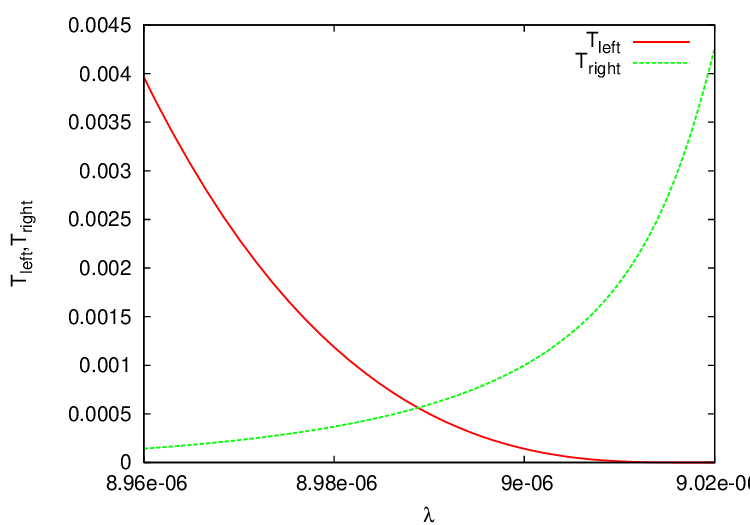}}}
\caption{(color online) Plot of the transmission coefficient $T$ vs anharmonicity strength $\lambda$.}
\end{figure}

With further increase in $\lambda$ parameter ($\sim 10^{-4}$) we now observe the condensate is tightly bound in the MSR by a high barrier on the left side and suffers tunneling through the right side barrier only.
\begin{figure}[hbpt]
\vspace{-10pt}
\centerline{
\hspace{-3.3mm}
\rotatebox{0}{\epsfxsize=8.8cm\epsfbox{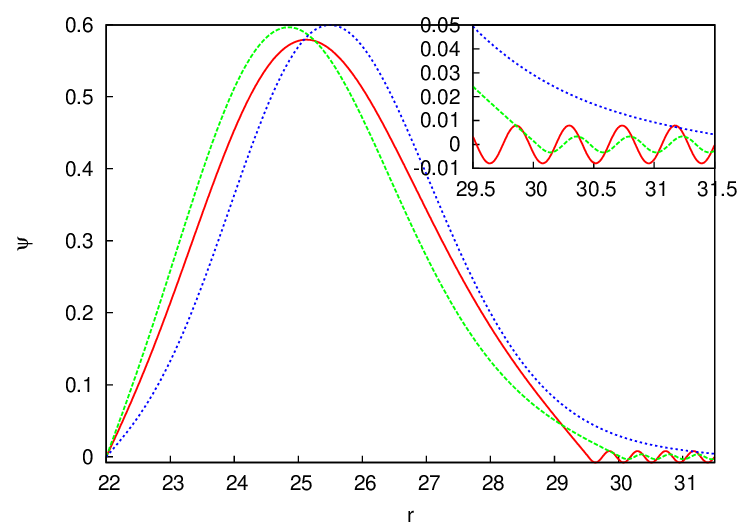}}}
\caption{(color online) plot of the condensate wave functions for various anharmonicity $\lambda$ and $N$=160. The blue dotted curve corresponds to $\lambda=9.8 \times 10^{-5}$, the green dashed curve and the red solid one correspond to $9.9 \times 10^{-5}$ and $1 \times 10^{-4}$ respectively.}
\end{figure}
Unlike the previous case of weak anharmonicity, 
here the MSR has no deep well on the right side. So after tunneling 
the condensate will behave as trapless uniform Bose gas. This feature 
is also clear from Fig. 9 where we observe that the condensate wave 
function is associated with oscillation on the right side. For better 
clarity the oscillatory part of the wave function is presented in the
inset of Fig. 9. It clearly shows that the condensate is basically confined 
in the MSR but a part leaks through the adjacent RIB. With increasing 
$\lambda$ the oscillation also increases. To get quantitative estimation of the MQT rate, in Table III we present the tunneling rate near the critical point. It again shows that MQT increases enormously near the critical point. In Fig.~10 we plot $\ln (\frac{S^{B}}{\hbar})$ as a function of $\ln (1-\frac{N}{N_{cr}})$ as before and fail to retrieve power law behavior.
\begin{center}
\begin{table}[!]
\caption{MQT rate for various $N$ near the $N_{cr}=165$ for $\lambda \sim 1 \times 10^{-4}$}
\begin{tabular}{|l|l|l|}
 \hline
N &  $\frac{S^{B}}{\hbar}$ & $\Gamma^{tunnel}_{N}$\\ \hline
160 & 0.980 & 318.566  \\ \hline
159 & 3.049 & 47.097  \\ \hline
158 & 5.108 & 6.493   \\ \hline
156 & 9.062 & 0.136   \\ \hline
\end{tabular}
\end{table}
\end{center}

\vspace*{.1cm}
\begin{figure}[hbpt]
\vspace{-10pt}
\centerline{
\hspace{-3.3mm}
\rotatebox{0}{\epsfxsize=8.8cm\epsfbox{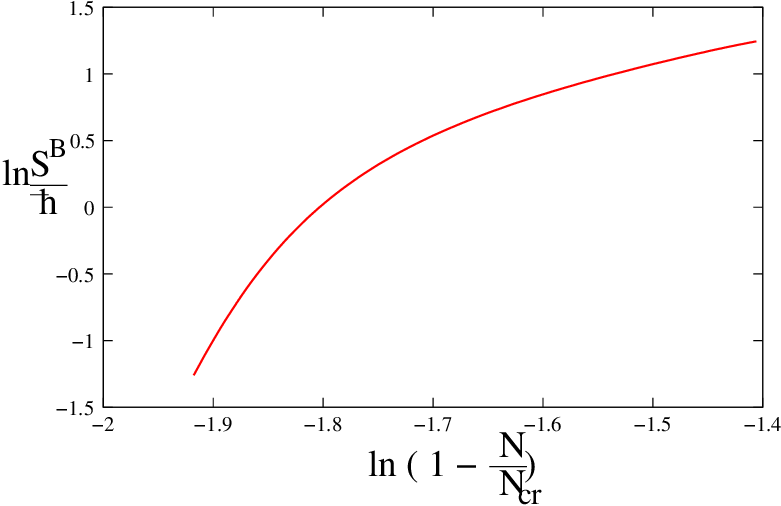}}}
\caption{(color online) Plot of $\ln (\frac{S^{B}}{\hbar})$ as a function of $\ln (1-\frac{N}{N_{cr}})$ for $\lambda \sim 10^{-4}$.}
\end{figure}

\section{Conclusion}
In our present work, we have studied the decay and tunneling of 
attractive Bose Einstein condensate in the finite trap by approximate 
but {\it ab-initio} many-body calculation. Main attention has been 
paid in the study of macroscopic quantum tunneling and the stability of attractive condensate when the 
effective trap height is tuned from very close to harmonic to weakly anharmonic. The use of correlated 
PH basis and the realistic van der Waals interaction correctly 
describes the many-body tunneling process and gives the realistic 
picture. The greater stability 
of attractive Bose gas in the weakly anharmonic trap is of special interest which significantlly differs from earlier mean-field results. In the present day experiments as the height of external confining potential is gradually reduced in a controlled fashion and quartic confinement of desired height is created, our theoretical results of MQT are also experimentally significant. We also observe that MQT is the significant decay mechanism in the anharmonic trap and there is a crucial interplay between the anharmonic strength and interatomic interaction. The deviation from the mean-field results in pure harmonic trap is attributed to two-body correlation and finite-range attraction of the realistic interatomic interaction. The possibility of dual tunneling for intermediate anharmonicity is also a new feature in quantum many-body calculation. Although there is no experimental evidence of the possibility of dual tunneling in such anharmonic trap, however with fine-controlled laser beam it may be possible in future to observe MQT through both the barriers. We also verify the power law behavior of MQT with $(1-\frac{N}{N_{cr}})$ near the criticality in pure harmonic trap. The results are in agreement with mean-field results. However we fail to retrieve any power law behavior in anharmonic trap.

\vskip 1cm
We thank Prof. Tapan Kumar Das and Dr. Anindya Biswas for valuable discussions. This work has been supported by a grant from the Department of Atomic 
Energy (DAE) (Grant No. 2009/37/23/BRNS/1903), Government of India,
and Department of Science and Technology (DST) 
[Fund No. SR/S2/CMP/0059(2007)], Government of 
India, under a research project. SKH acknowledges the 
Council of Scientific and Industrial Research (CSIR), India for the Senior 
Research Fellowship. \\

\end{document}